\documentclass[12pt,superscriptaddress]{article}
\usepackage{palatino,cite,epsfig,amsmath,amssymb}

\allowdisplaybreaks

\usepackage{psfrag}
\newcommand{\cha}{\tilde{\chi}}
\newcommand{\neu}{\tilde{\chi}^0}
\def\SPHENO{SPheno\,2.2.0}

\begin{document}
\begin{flushright}
CERN-PH-TH/2004-044 \\
DESY 04-044 \\
HEPHY-PUB 786/04 \\
LAPTH-1032/04 \\
ZU-TH 04/04
\end{flushright}
%\vspace{1cm}

\begin{center}

{\Large\bf 
  Reconstructing Supersymmetric Theories \\[2mm]
  by Coherent LHC / LC Analyses%
}
 
\vspace{1cm}

{\sc 
B.~C.~Allanach$^{1}$, G.~A.~Blair$^{2,3}$, 
S.~Kraml$^{4,5}$, H.-U.~Martyn$^6$,\\[2mm] 
G.~Polesello$^{7}$, W.~Porod$^{8}$, 
and P.~M.~Zerwas$^{2}$
}

\vspace*{1cm}

{\sl
$^1$ LAPTH, Annecy-le-Vieux, France.\\[4mm]
$^2$Deutches Elektronen-Synchrotron DESY, D-22603 Hamburg, Germany.\\[4mm]
$^3$Royal Holloway University of London, Egham, Surrey. TW20 0EX, UK.\\[4mm]
$^4$Inst. f. Hochenergiephysik, \"Osterr. Akademie d. Wissenschaften, 
    Vienna, Austria.\\[4mm]
$^5$CERN, Dept. of Physics, TH Division, Geneva, Switzerland.\\[4mm]
$^6$I. Physik. Institut, RWTH Aachen, D-52074 Aachen, Germany.\\[4mm]
$^7$INFN, Sezione di Pavia, Via Bassi 6, Pavia 27100, Italy.\\[4mm]
$^8$Institut f\"ur Theoretische Physik, Universit\"at Z\"urich, Switzerland.
}

\end{center}

\vspace*{1cm}

\begin{abstract}
Supersymmetry analyses will potentially be a central
area for experiments at the LHC and at a future $e^+ e^-$ linear collider.
Results from the two facilities will mutually complement and
augment each other so that
a comprehensive 
and precise picture of the supersymmetric world can be developed.
We will demonstrate in this report how coherent analyses at LHC and LC
experiments can be used to explore the breaking mechanism of supersymmetry
and to reconstruct the fundamental theory at high energies, in particular
at the grand unification scale. This will be exemplified for minimal
supergravity in detailed experimental simulations performed
for the Snowmass reference point SPS1a.

\end{abstract}

\newpage
\section{Physics Base}

The roots of standard particle physics are expected to go as 
deep as the Planck length of $10^{-33}$~cm where gravity is 
intimately linked to the particle system.  A stable bridge
between the electroweak energy scale of 100~GeV and
the vastly different Planck scale of $\Lambda_{\rm PL}\sim10^{19}$~GeV,
and the (nearby) grand unification scale $\Lambda_{\rm GUT}\sim10^{16}$~GeV,
is provided by supersymmetry.  If this scenario is realized in nature,
experimental methods must be developed to shed light on the physics 
phenomena near $\Lambda_{\rm GUT}$/$\Lambda_{\rm PL}$.  Among other
potential tools, the extrapolation of supersymmetry (SUSY) 
parameters measured at 
the LHC and an e$^+$e$^-$ linear collider with high precision, can play
a central r\^ole~\cite{r1}.  A rich ensemble of gauge and Yukawa couplings,
and of gaugino/higgsino and scalar particle
masses allows the detailed study of the 
supersymmetry breaking mechanism and the reconstruction of the 
physics scenario near the GUT/PL scale.

The reconstruction of physical structures at energies more than fourteen
orders above the energies available through accelerators is
a demanding task.  Not only must a comprehensive picture
be delineated near the electroweak scale, but
the picture must be drawn, moreover, as precisely as possible to keep
the errors small enough so that they do not blow up beyond control when the
SUSY parameters are extrapolated over many orders of magnitude.
The LHC~\cite{r2} and a future e$^+$e$^-$ linear collider (LC)~\cite{r3}
are a perfect tandem for solving such a problem: {\bf (i)} While the
colored supersymmetric particles, gluinos and squarks, can be
generated with large rates for masses up to 
2 to 3~TeV at the LHC, the strength of e$^+$e$^-$ linear colliders
is the comprehensive coverage of the non-colored particles, 
charginos/neutralinos and sleptons.  If the extended Higgs spectrum
is light, the Higgs particles can be discovered and investigated at
both facilities; heavy Higgs bosons can be produced at the 
LHC in a major part of the parameter space; at an e$^+$e$^-$
collider, without restriction, for masses up to the beam energy;
{\bf (ii)} If the analyses are performed coherently, the 
accuracies in measurements of  cascade decays at LHC
and in threshold production as well as decays of supersymmetric
particles at LC complement and augment
each other {\it mutually} so that a high-precision
picture of the supersymmetric parameters at the electroweak scale
can be drawn.  Such a comprehensive and precise picture is 
necessary in order to carry out the evolution of the 
supersymmetric parameters to high scales, driven by perturbative 
loop effects that involve the entire supersymmetric particle spectrum.

Minimal supergravity (mSUGRA) provides us with a scenario within
which these general ideas can be quantified.  The  form of this
theory has been developed in great detail, creating a
platform on which semi-realistic experimental studies can be performed.
Supersymmetry is broken in mSUGRA in a hidden sector and the breaking
is transmitted to our eigenworld by gravity~\cite{Chamseddine:jx}.  This
mechanism suggests, yet does not enforce, the universality of the
soft SUSY breaking parameters -- gaugino and scalar masses, and
trilinear couplings -- at a scale that is generally identified with
the unification scale.  The (relatively) small number of
parameters renders mSUGRA a well-constrained system that suggests itself
in a natural way as a test ground for coherent experimental analyses
at LHC and LC.  The procedure will be exemplified for a specific set of
parameters, defined as SPS1a among the Snowmass reference points~\cite{r4}.

\section{Minimal Supergravity: SPS1a}

The mSUGRA Snowmass reference point SPS1a is characterised by
the following values~\cite{r4}: 
\begin{eqnarray}
  \begin{array}{ll}
    M_{1/2} = 250~{\rm GeV} \qquad & M_0=100~{\rm GeV} \\
    A_0=-100~{\rm GeV} & {\rm sign}(\mu)=+\\
    \tan\beta=10 & \\
  \end{array}
\end{eqnarray}

\noindent
for the universal gaugino mass $M_{1/2}$, 
the scalar mass $M_0$, the trilinear coupling
$A_0$, the sign of the higgsino parameter $\mu$, and $\tan\beta$,
the ratio of the vacuum-expectation values of the two Higgs fields.
As the modulus of the Higgsino parameter is fixed at the
electroweak scale by requiring
radiative electroweak symmetry breaking, $\mu$ is finally given by:
\begin{equation}
  \mu = 357.4~{\rm GeV} 
\end{equation}

This reference point is compatible with the constraints
from low-energy precision data, predicting 
BR($b\rightarrow s\gamma) =2.7\cdot 10^{-4}$
and $\Delta[g_\mu-2]= 17\cdot 10^{-10}$.  The amount of cold dark matter
is, with $\Omega_\chi h^2=0.18$, on the high side but still compatible
with recent WMAP data \cite{r5} if evaluated on their own without 
reference to other experimental results;
moreover, only a slight shift in $M_0$ downwards drives the value
to the central band of the data while such a shift does not alter
any of the conclusions in this report in a significant way.

\begin{figure}[t]
\begin{center}
\setlength{\unitlength}{1mm}
\begin{picture}(75,85)
\psfrag{h}{\Huge $h^0$}
\psfrag{J}{\Huge $H^0$}
\psfrag{Z}{\Huge $\,A^0$}
\psfrag{I}{\Huge $H^\pm$}
\psfrag{N}{\Huge $\neu_1$}
\psfrag{M}{\Huge $\neu_2$}
\psfrag{O}{\Huge $\cha^\pm_1$}
\psfrag{P}{\Huge $\tilde g$}
\psfrag{Y}{\Huge $\neu_3$}
\psfrag{X}{\Huge $\tilde{\tau}_1$}
\psfrag{T}{\Huge $\tilde{\tau}_2$}
\psfrag{R}{\Huge $\tilde{e}_{\rm R}$}
\psfrag{L}{\Huge $\tilde{e}_{\rm L}$}
\psfrag{V}{\Huge $\tilde{\nu}_\tau$}
\psfrag{U}{\Huge $\tilde{\nu}_e$}
\psfrag{B}{\Huge $\tilde{t}_1$}
\psfrag{A}{\Huge $\tilde{b}_1$}
\psfrag{C}[r][r]{\Huge $\tilde{q}_{\rm R}, \tilde{b}_2$}
\psfrag{D}{\Huge $\tilde{q}_{\rm L}$}
\psfrag{F}[r][r]{\Huge $\tilde{t}_2$}
\psfrag{Q}[r][r]{\Huge $\neu_4, \cha^\pm_2\!\!\!\!\!\!$}
\put(0,0){%
\epsfig{figure=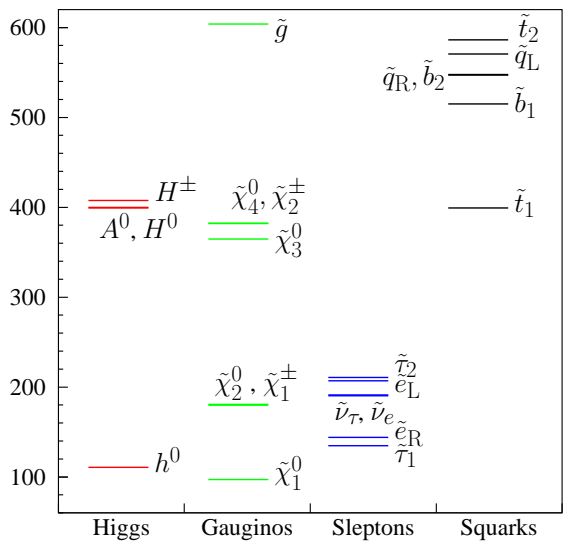, angle=0, width=7.5cm,
        viewport=32 0 540 505, clip=true}
}
\end{picture}
\end{center}
\vspace{-5mm}
\caption{\it Spectrum of Higgs, gaugino/higgsino and sparticle masses
    in the mSUGRA scenario SPS1a; masses in {\normalfont GeV}.}
\label{fig:SPS1a_spectrum}
\end{figure}

The form of the supersymmetric mass spectrum of SPS1a is shown
in Fig.~\ref{fig:SPS1a_spectrum}.  In this scenario the squarks and gluinos
can be studied very well at the LHC while the non-colored gauginos
and sleptons can be analyzed partly at LHC and in comprehensive form
at an e$^+$e$^-$ linear collider operating at a total energy below
1 TeV with high integrated luminosity close to 1~ab$^{-1}$.

%\vskip 1 cm

The masses can best be obtained at {\underline {LHC}} by analyzing 
edge effects in the
cascade decay spectra. The basic starting point is the identification
of the a sequence of two-body decays:
\mbox{$\tilde q_L\rightarrow\tilde\chi^0_2 q\rightarrow\tilde\ell_R\ell q
\rightarrow \tilde\chi^0_1\ell\ell q$}. This is effected  through the 
detection of an edge structure of the invariant mass 
of opposite-sign same-flavour leptons
from the $\chi^0_2$ decay in events with multi-jets and $E_T^{miss}$.
One can then measure the kinematic edges 
of the invariant mass distributions among the two leptons and the jet resulting
from the above chain, and 
thus an approximately model-independent determination 
of the masses of the involved sparticles is obtained. 
This technique was developed in Refs.~\cite{HinPai, Cambr} and is worked 
out in detail for point SPS1a in Ref.~\cite{SPSLHC}.
The four sparticle masses ($\tilde q_L$, $\tilde\chi^0_2$,
$\tilde\ell_R$, and $\tilde\chi^0_1$) thus measured are used as an input
to additional analyses which rely on the knowledge
of the masses of the lighter gauginos in order 
to extract masses from the observed 
kinematic structures. Examples are the studies of 
the decay 
\mbox{$\tilde g\rightarrow\tilde b_1 b\rightarrow \tilde\chi^0_2 bb$},
where the reconstruction of the gluino and sbottom 
mass peaks relies on an approximate 
full reconstruction of the $\tilde\chi^0_2$, 
and the shorter decay chains \mbox{$\tilde q_R\rightarrow q \tilde\chi^0_1$}
and \mbox{$\tilde\chi^0_4\rightarrow\tilde\ell\ell$}, 
which require the 
knowledge of the sparticle masses downstream of the cascade.
For SPS1a the heavy Higgs bosons can also be searched for in 
the decay chain: 
$A^0 \to \tilde \chi^0_2 \tilde \chi^0_2
     \to \tilde \chi^0_1 \tilde \chi^0_1 l^+ l^- l^+ l^-$ 
\cite{moortgat}. The invariant four-lepton mass depends
sensitively on $m_{A^0}$ and $m_{\chi^0_1}$. The same holds true for $H^0$.
Note however that the main source of the neutralino final states are
$A^0$ decays, and that the two Higgs bosons $A^0$ and $H^0$ cannot 
be discriminated in this channel. 

The mass measurements obtained at the LHC are thus very 
correlated among themselves, and this correlation must be taken into
account in the fitting procedure.  Another source of correlation 
comes from the fact that in most cases the uncertainty on the
mass measurement is dominated by the systematic uncertainty on the 
hadronic energy scale of the experiment, which will affect 
all the measurements involving jets approximately by the same 
amount and in the same direction.

\begin{figure*}
\setlength{\unitlength}{1mm}
\begin{center}
\begin{picture}(160,170)
\put(-3,162){\mbox{\bf a) 
    \quad\quad\quad\quad\quad\quad\quad \large LHC}}
\put(-3,80){\mbox{\epsfig{figure=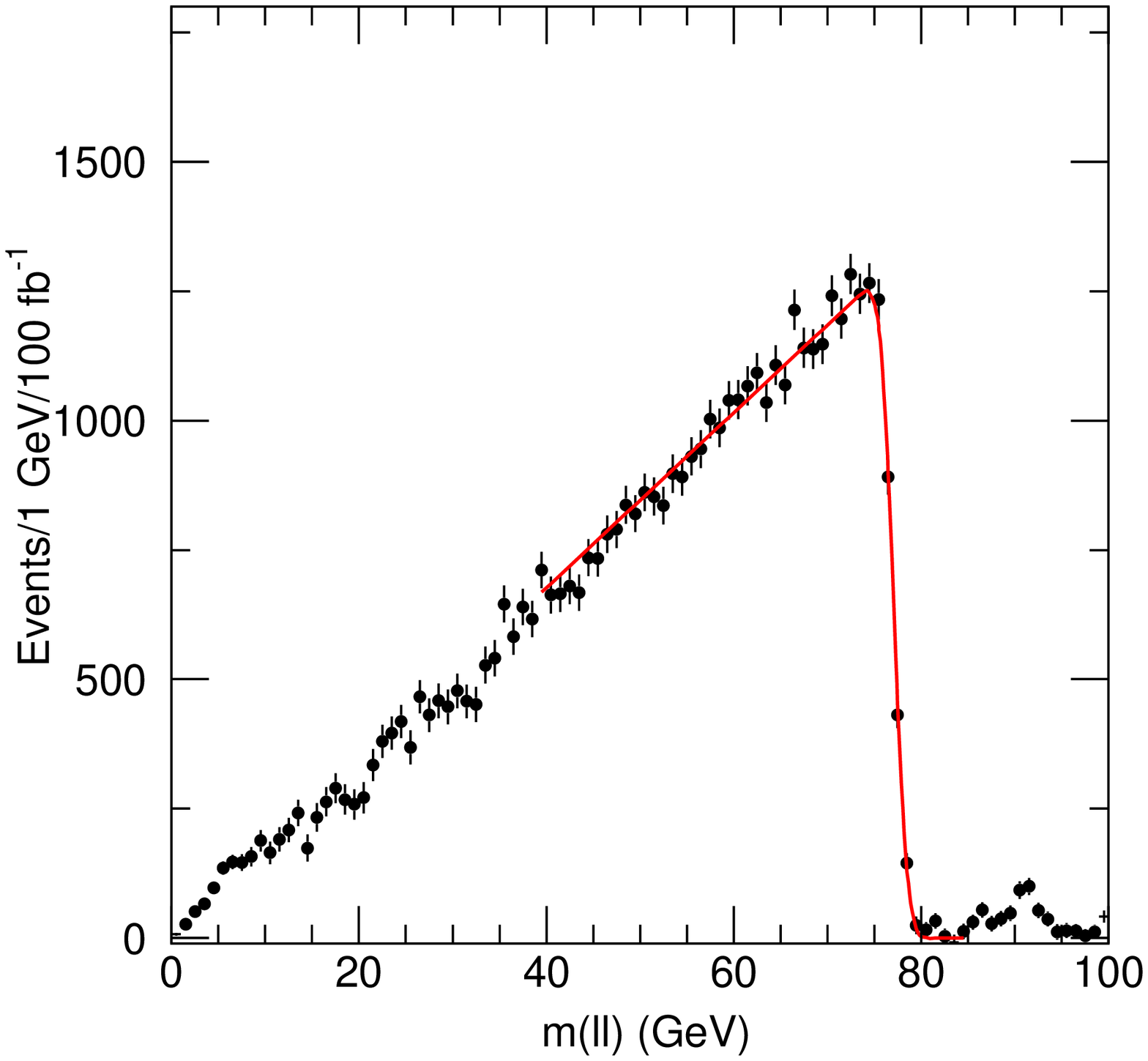,
                                   height=8cm,width=8cm,angle=0}}}
\put(-3,-3){\mbox{\epsfig{figure=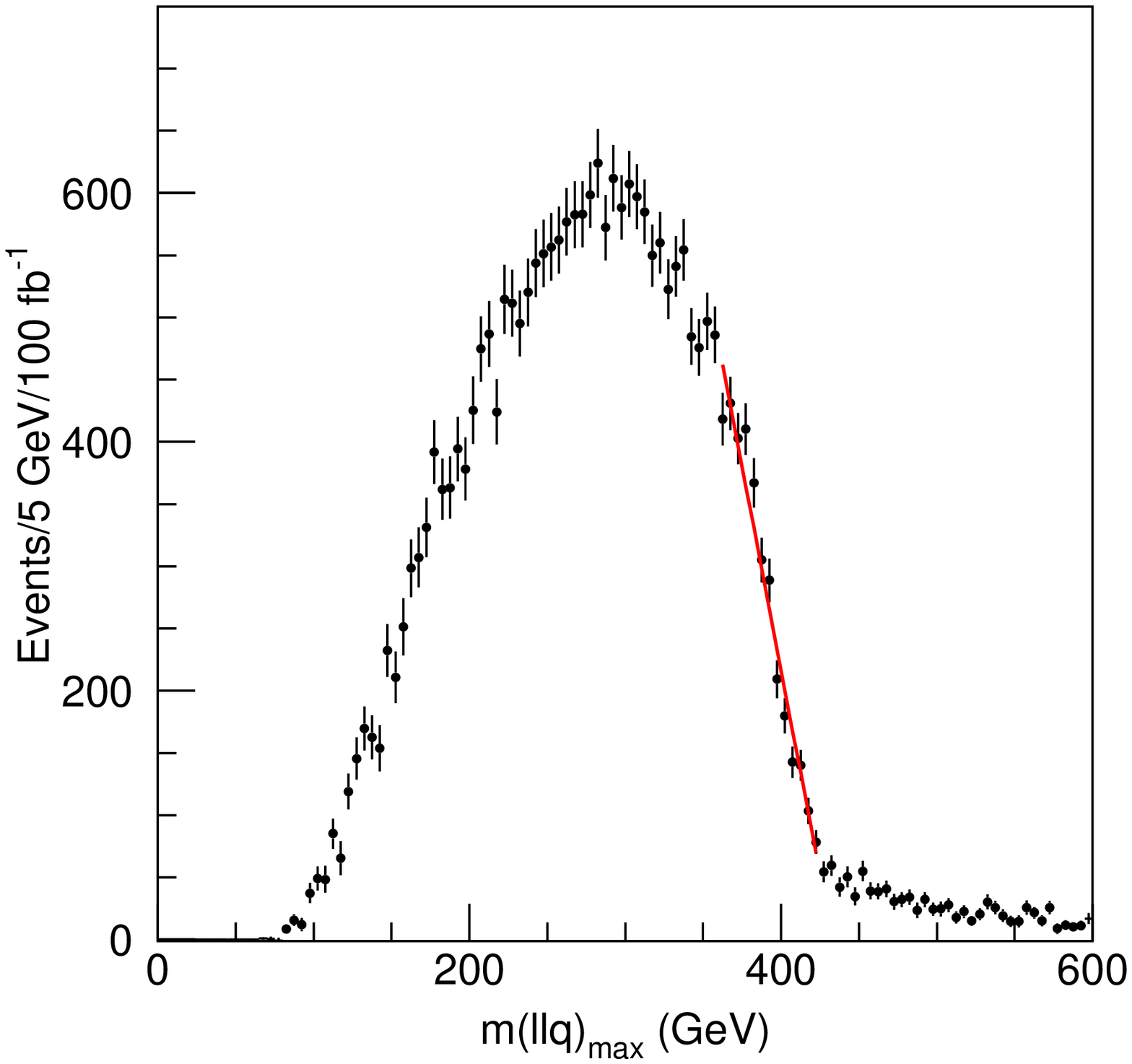,
                                   height=8cm,width=8cm,angle=0}}}
\put(84,162){\mbox{\bf b) 
    \quad\quad\quad\quad\quad\quad\quad \large LC}}
\put(81,84.5){\mbox{\epsfig{
              figure=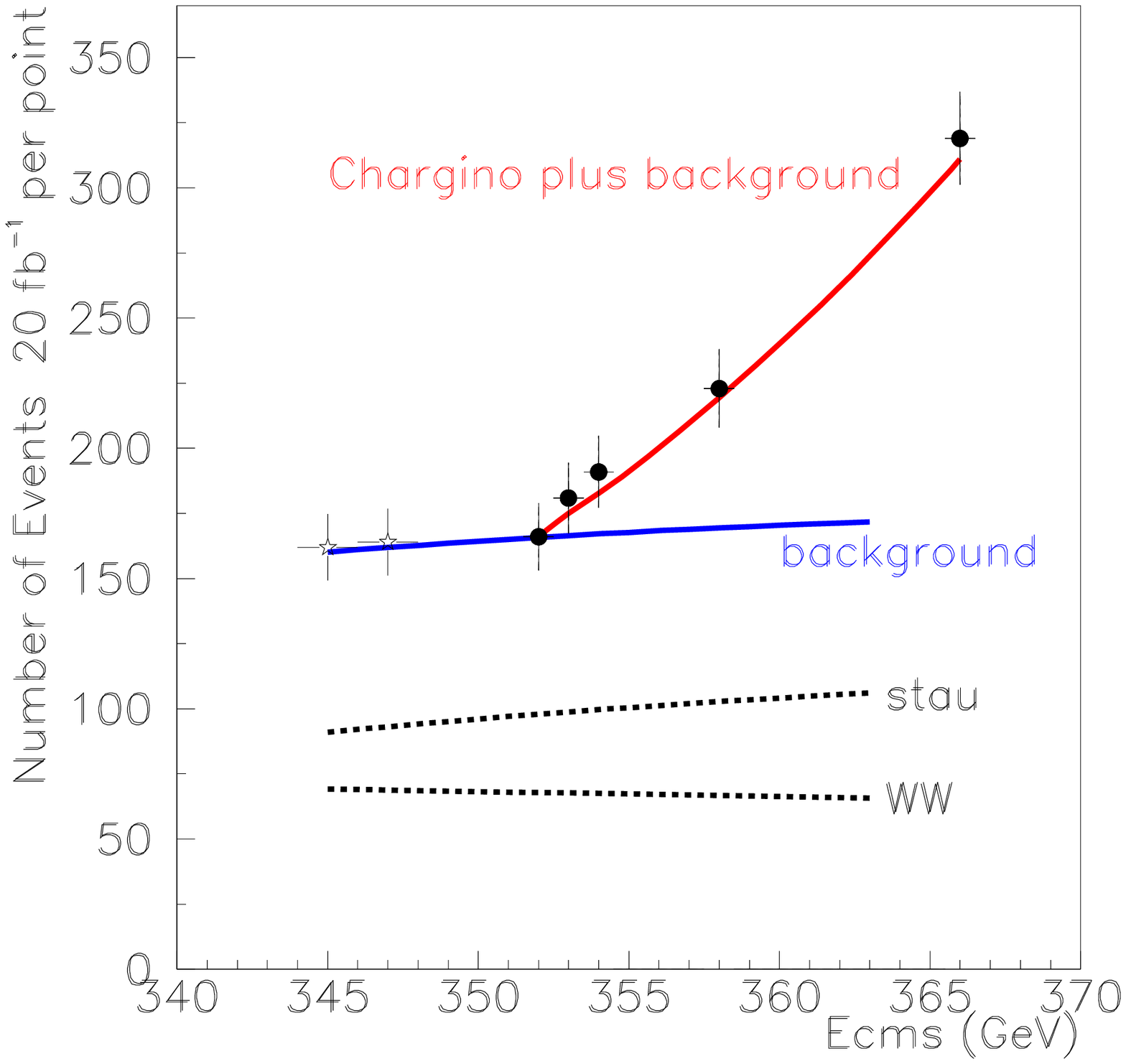,width=8.45cm,height=7.5cm}}}
\put(98,156){\mbox{$e^+_R e^-_L \to \tilde\chi^+_1 \tilde\chi^-_1 $}}
\put(80,0){\mbox{\epsfig{figure=sps1_ser_energy.eps,height=17.cm}}}
\put(98,72){\mbox{$e^+_L e^-_R \to \tilde e_R \tilde e_R
                    \to e^+\tilde\chi^0_1\,  e^-\tilde\chi^0_1 $} }
\end{picture}
\end{center}
\caption {{\it {\bf a)
Examples of LHC edge spectra.} 
Upper: Two-lepton mass with kinematic edge from the $\tilde \chi^0_2$ decay.
Lower: Kinematic end-point involving the mass spectrum of two leptons 
       and a jet for $\tilde q_L$ decay.
{\bf  b) 
Examples of LC analyses.} 
Upper: Threshold scan for chargino production
       $e^+_R e^-_L \to \tilde\chi^+_1 \tilde\chi^-_1 $
       including the backgrounds 
       from $W^+W^-$ and $\tilde\tau^+_1 \tilde\tau^-_1$;
       observed event rates correspond to the integrated luminosity 
       ${\cal L} = 100~{\rm fb}^{-1}$.
Lower: Electron decay spectrum of the continuum reaction
       $e^+_L e^-_R \to \tilde e_R^+ \tilde e_R^-
       \to e^+\tilde\chi^0_1  e^-\tilde\chi^0_1 $,
       assuming $\sqrt{s} = 400$~GeV and ${\cal L} = 200~{\rm fb}^{-1}$.
}}
\label{fig:exp_plots}
\end{figure*}

Two characteristic examples are shown in  Fig.~\ref{fig:exp_plots}a: 
the edge in the di-lepton invariant mass distribution 
for the decay $\tilde\chi^0_2\rightarrow\tilde\ell_R\ell$,
and the endpoint of the invariant mass of two leptons plus a jet
for the decay $\tilde q_L\rightarrow\tilde\chi^0_2 q$.

At {\underline {linear colliders}} very precise mass values can be extracted 
from decay spectra and threshold scans \cite{r6A,r6B,r6C}.
The excitation curves for chargino production
in S-waves~\cite{r7} rise steeply with the velocity of the particles
near the threshold and thus are very sensitive to their mass values;
the same is true for mixed-chiral selectron pairs in
$e^+e^-\to \tilde e_R^+ \tilde e_L^-$ 
and for diagonal pairs in 
$e^-e^-\to \tilde e_R^- \tilde e_R^-, \;  \tilde e_L^- \tilde e_L^-$
collisions \cite{r6C}.  
Other scalar sfermions, as well as neutralinos, 
are produced generally in P-waves, with a
somewhat less steep threshold behaviour proportional to the
third power of the velocity~\cite{r6C,r8}.  Additional information,
in particular on the lightest neutralino $\tilde{\chi}^0_1$, can
be obtained from decay spectra.  
Two characteristic examples
of a threshold excitation curve and a decay spectrum
are depicted in Fig.~\ref{fig:exp_plots}b.

Typical mass parameters and the related measurement
errors are presented in Table~\ref{tab:massesA}.  
The column denoted ``LHC'' collects
the errors from the LHC analysis, the column ``LC'' the errors
expected from the LC operating at energies up to 1~TeV with an
integrated luminosity of $\sim1$~ab$^{-1}$.  
The error estimates are based on detector simulations for the production
of the light sleptons, $\tilde e_R$, $\tilde \mu_R$ and $\tilde \tau_1$,
in the continuum.
For the light neutralinos and the light chargino threshold scans have 
been simulated. 
Details will be given elsewhere; see also Ref.\cite{R13A}.
The expected precision of the other particle masses is taken from
Ref.\cite{r6C},
or it is obtained by scaling the LC errors from the previous 
analysis in Ref.\cite{r6A},
taking into account the fact that the 
$\tilde \chi^0 / \tilde \chi^\pm$ cascade decays
proceed dominantly via $\tau$ leptons in the reference point SPS1a,
which is experimentally  challenging.
The third column of Tab.~\ref{tab:massesA} denoted 
``LHC+LC'' presents the corresponding errors if the
experimental analyses are performed coherently, i.e. the
light particle spectrum, studied at LC with very high precision,
is used as an input set for the LHC analysis.

\renewcommand{\arraystretch}{1.1}
\begin{table}
\begin{center}
\begin{tabular}{|c||c||c|c||c|}
\hline
               & Mass, ideal & ``LHC''  &\ ``LC''\ & ``LHC+LC''
\\ \hline\hline
$\tilde{\chi}^\pm_1$ & 179.7 &          & 0.55   &  0.55  \\
$\tilde{\chi}^\pm_2$ & 382.3 &     --   & 3.0    &  3.0   \\
$\tilde{\chi}^0_1$   &  97.2 &     4.8  & 0.05   &  0.05  \\
$\tilde{\chi}^0_2$   & 180.7 &     4.7  & 1.2    &  0.08  \\
$\tilde{\chi}^0_3$   & 364.7 &          & 3-5    &  3-5   \\
$\tilde{\chi}^0_4$   & 381.9 &     5.1  & 3-5    &  2.23  \\
\hline
$\tilde{e}_R$        & 143.9 &     4.8  & 0.05   &  0.05  \\
$\tilde{e}_L$        & 207.1 &     5.0  & 0.2    &  0.2   \\
$\tilde{\nu}_e$      & 191.3 &     --   & 1.2    &  1.2   \\
$\tilde{\mu}_R$      & 143.9 &     4.8  & 0.2    &  0.2  \\
$\tilde{\mu}_L$      & 207.1 &     5.0  & 0.5    &  0.5  \\
$\tilde{\nu}_\mu$    & 191.3 &     --   &        &       \\
$\tilde{\tau}_1$     & 134.8 &     5-8  & 0.3    &  0.3  \\
$\tilde{\tau}_2$     & 210.7 &     --   & 1.1    &  1.1  \\
$\tilde{\nu}_\tau$   & 190.4 &     --   & --     &  --   \\
\hline
$\tilde{q}_R$        & 547.6 &    7-12  &    --  & 5-11  \\
$\tilde{q}_L$        & 570.6 &     8.7  &    --  &  4.9  \\
$\tilde{t}_1$        & 399.5 &          & 2.0    &  2.0  \\
$\tilde{t}_2$        & 586.3 &          &   --   &       \\
$\tilde{b}_1$        & 515.1 &     7.5  &    --  &  5.7  \\
$\tilde{b}_2$        & 547.1 &     7.9  &    --  &  6.2  \\
\hline
$\tilde{g}$          & 604.0 &     8.0  &    --  &  6.5  \\
\hline
$h^0$                & 110.8 &     0.25 & 0.05   & 0.05   \\
$H^0$                & 399.8 &          & 1.5    & 1.5   \\
$A^0$                & 399.4 &          & 1.5    & 1.5   \\
$H^{\pm}$            & 407.7 &     --   & 1.5    & 1.5   \\\hline 
\end{tabular}\\
\end{center}
\caption {{\it Accuracies for representative mass measurements
at ``LHC'' and ``LC'', and in coherent ``LHC+LC'' analyses
for the reference point SPS1a [masses in {\rm GeV}].
$\tilde q_L$ and $\tilde q_R$ represent the flavours
$q=u,d,c,s$ which cannot be distinguished at LHC.
Positions marked by bars cannot be filled either due to 
kinematical restrictions or due to small signal rates; blank positions 
could eventually be filled after significantly more investments 
in experimental simulation efforts than performed until now.
The ``LHC'' and ``LC'' errors have been derived in Ref.~\cite{SPSLHC}
and Ref.~\cite{LC-errors}, respectively, in this document.}} 
\label{tab:massesA}
\end{table}
 
Mixing parameters must be obtained from measurements of cross
sections and polarization asymmetries, 
in particular from the production of chargino pairs and
neutralino pairs~\cite{r7,r8}, both in diagonal or mixed form: 
$e^+e^- \rightarrow {\tilde{\chi}^+_i}{\tilde{\chi}^-_j}$
[$i$,$j$ = 1,2] and ${\tilde{\chi}^0_i} {\tilde{\chi}^0_j}$ 
[$i$,$j$ = 1,$\dots$,4]. 
The production cross sections for
charginos are binomials of $\cos\,2\phi_{L,R}$, the mixing angles
rotating current to mass eigenstates. Using polarized electron
and positron beams, the cosines can be determined in a model-independent
way.

\renewcommand{\arraystretch}{1.1}
\begin{table}[t]
\begin{center}
\begin{tabular}{|c||c|c|}
\hline
           & Parameter, ideal   & {``LHC+LC''} errors
\\ \hline\hline
 $M_1$        & 101.66   &   0.08  \\
 $M_2$        & 191.76   &   0.25  \\
 $M_3$        & 584.9    &   3.9   \\
\hline
$\mu$         & 357.4    &   1.3   \\
\hline  
 $M^2_{L_1}$  &$3.8191 \cdot 10^4$ & 82.   \\
 $M^2_{E_1}$  &$1.8441 \cdot 10^4$ & 15.   \\
 $M^2_{Q_1}$  &$29.67 \cdot 10^4$  & $0.32\cdot 10^4$ \\
 $M^2_{U_1}$  &$27.67 \cdot 10^4$ &  $0.86 \cdot 10^4$ \\
  $M^2_{D_1}$ &$27.45 \cdot 10^4$ &  $0.80 \cdot 10^4$ \\
 $M^2_{L_3}$  &$3.7870 \cdot 10^4$&  360.  \\
 $M^2_{E_3}$  &$1.7788 \cdot 10^4$&   95.   \\
 $M^2_{Q_3}$  &$24.60 \cdot 10^4$ & $0.16 \cdot 10^4$\\
 $M^2_{U_3}$  &$17.61 \cdot 10^4$ & $0.12 \cdot 10^4$\\
 $M^2_{D_3}$  &$27.11 \cdot 10^4$ & $0.66 \cdot 10^4$\\
\hline
$M^2_{H_1} $  & \hspace*{5mm}$3.25 \cdot 10^4$ & $ 0.12 \cdot 10^4$ \\
$M^2_{H_2} $  &$-12.78 \cdot 10^4$& $0.11 \cdot 10^4$  \\
$A_t $        & $-497.$       &  9.    \\
\hline
$\tan\beta$   & 10.0              &  0.4   \\
\hline
\end{tabular}\\
\end{center}
\caption {{\it The extracted SUSY Lagrange mass and Higgs parameters 
at the electroweak scale in the reference point SPS1a
[mass units in {\rm GeV}].
}} 
\label{tab:params}
\end{table}

Based on this high-precision information, the fundamental SUSY
parameters can be extracted at low energy in analytic form. To lowest
order:
\begin{eqnarray}
\left|\mu\right|&=&M_W[\Sigma + \Delta[\cos2\phi_R+\cos2\phi_L]]^{1/2}
\nonumber\\
\mbox{sign}(\mu)&= &[ \Delta^2
                   -(M^2_2-\mu^2)^2-4m^2_W(M^2_2+\mu^2) \nonumber \\
 & &                   -4m^4_W\cos^2 2\beta]/8 m_W^2M_2|\mu|\sin2\beta 
\nonumber\\
M_2&=&M_W[\Sigma - \Delta(\cos2\phi_R+\cos2\phi_L)]^{1/2}\nonumber\\
|M_1|&=& \left[ \textstyle \sum_i m^2_{\tilde{\chi}_i^0}  
                 -M^2_2-\mu^2-2M^2_Z\right]^{1/2}
\nonumber\\
|M_3|&=&m_{\tilde{g}} \nonumber\\
\tan\beta&=&\left[\frac{1+\Delta (\cos 2\phi_R-\cos 2\phi_L)}
           {1-\Delta (\cos 2\phi_R-\cos 2\phi_L)}\right]^{1/2} 
\label{eqn:basicLE}
\end{eqnarray}
where $\Delta = (m^2_{\tilde{\chi}^\pm_2}-m^2_{\tilde{\chi}^\pm_1})/(4M^2_W)$
and 
$\Sigma =  (m^2_{\tilde{\chi}^\pm_2}+m^2_{\tilde{\chi}^\pm_1})/(2M^2_W) -1$.
The signs of $M_{1,3}$ with respect to $M_2$ follow from measurements of
the cross sections for ${\tilde{\chi}} {\tilde{\chi}}$ production and 
gluino processes. In practice one-loop corrections to the mass relations 
have been used to improve on the accuracy.  
 
The mass parameters of 
the sfermions are directly related to the
physical masses if mixing effects are negligible: 
\begin{equation}
m^2_{\tilde{f}_{L,R}}=M^2_{L,R}+m^2_f + D_{L,R} 
\end{equation}
with $D_{L} = (T_3 - e_f \sin^2 \theta_W) \cos 2 \beta \, m^2_Z$ 
and $D_{R} = e_f \sin^2 \theta_W \cos 2 \beta \, m^2_Z$ 
denoting the D-terms.  The non-trivial
mixing angles in the sfermion sector of the third generation
can be measured in a way similar to the charginos and neutralinos.  
The sfermion production cross sections for
longitudinally polarized e$^+$/e$^-$ beams are bilinear
in $\cos$/$\sin2\theta_{\tilde f}$.  The mixing angles and the two physical
sfermion masses are related to the tri-linear couplings $A_f$,
the higgsino mass parameter $\mu$ and $\tan\beta(\cot\beta)$
for down(up) type sfermions by:
\begin{equation}
A_f-\mu\tan\beta(\cot\beta)=\frac{m^2_{\tilde{f}_1}-m^2_{\tilde{f}_2}}{2 m_f}\sin2\theta_{\tilde f} ~~~~~[f:{\rm down(up)~type}]
\end{equation}
This relation gives us the opportunity to measure 
$A_f$ if $\mu$ has been determined
in the chargino sector.

Accuracies expected for the SUSY Lagrange parameters at the
electroweak scale for the reference point SPS1a are shown in
Table~\ref{tab:params}.  The errors are presented for 
the coherent ``LHC+LC'' analysis.  They have been obtained by fitting
the LHC observables and the masses of SUSY particles and Higgs bosons
accessible at a 1 TeV Linear Collider.
For the fit the programs SPheno2.2.0 \cite{Porod:2003um} and 
MINUIT96.03 \cite{James:1975dr} have been used.
The electroweak gaugino and higgs/higgsino parameters cannot be
determined individually through mass measurements at the LHC
as the limited number of observable masses leaves this sector 
in the SPS1a system under-constrained. Moreover, the Lagrange 
mass parameters in the squark sector can be determined 
from the physical squark masses with sufficient accuracy only
after the LHC mass measurements are complemented by LC measurements
in the chargino/neutralino sector; this information is necessary as 
the relation between the mass parameters is affected by large loop 
corrections.

\section{Reconstruction of the Fundamental SUSY Theory}

As summarized in the previous section, 
the minimal supergravity scenario mSUGRA is characterized
by the universal gaugino parameter $M_{1/2}$, the scalar
mass parameter $M_0$ and the trilinear coupling $A_0$, all
defined at the grand unification scale.  These parameters
are complemented by the sign of the higgs/higgsino mixing parameter
$\mu$, with the modulus determined by radiative symmetry
breaking, and the mixing angle, $\tan\beta$, in the Higgs sector.

The fundamental mSUGRA parameters at the GUT scale are
related to the low-energy parameters at the electroweak scale
by supersymmetric renormalization group 
transformations (RG)~\cite{RGE1,RGE2}
which to leading order generate the evolution for
\begin{center}
\begin{tabular}{lclr}
 gauge couplings &:&  $\alpha_i = Z_i \, \alpha_U$ & (5) \\
  gaugino mass parameters &:& $M_i = Z_i \, M_{1/2}$ & (6) \\
 scalar mass parameters &:&   $M^2_{\tilde\jmath} = M^2_0 + c_j M^2_{1/2} +
        \sum_{\beta=1}^2 c'_{j \beta} \Delta M^2_\beta$  & (7) \\
  trilinear  couplings &:&  $A_k = d_k A_0   + d'_k M_{1/2}$ & (8) 
\end{tabular}
\end{center}
\refstepcounter{equation}
\refstepcounter{equation}
\label{eq:gaugino}
\refstepcounter{equation}
\label{eq:squark} 
\refstepcounter{equation}
The index $i$ runs over the gauge groups $i=SU(3)$, $SU(2)$, $U(1)$.
To leading order, the gauge couplings, and the gaugino and scalar mass
parameters of soft--supersymmetry breaking depend on the $Z$ transporters
with 
\begin{eqnarray}
Z_i^{-1} =  1 + b_i \frac{\alpha_U}{ 4 \pi}
             \log\left(\frac{M_U}{ M_Z}\right)^2 
\end{eqnarray}
and $b[SU_3, SU_2, U_1] = -3, \, 1, \, 33 / 5$;
the scalar mass parameters depend 
also on the Yukawa couplings $h_t$, $h_b$, $h_\tau$
of the
top quark, bottom quark and $\tau$ lepton.
The coefficients $c_j$ [$j=L_l, E_l, Q_l, U_l, D_l, H_{1,2}$; $l=1,2,3$] 
for the slepton and squark doublets/singlets of generation $l$, 
and for the two Higgs doublets
are linear combinations of the evolution
coefficients $Z$; the coefficients $c'_{j \beta}$ are of order unity. 
The shifts $\Delta M^2_\beta$ are nearly zero for the first two families of 
sfermions but they can be rather large for the third family and for the 
Higgs mass
parameters, depending on the coefficients $Z$, 
the universal parameters $M^2_0$, $M_{1/2}$ and $A_0$,
and on the Yukawa couplings $h_t$, $h_b$, $h_\tau$.
The coefficients $d_k$ of the trilinear
couplings $A_k$ [$k=t,b,\tau$]  
depend on the corresponding Yukawa couplings 
and they are approximately unity for the
first two generations while being O($10^{-1}$) 
and smaller if the Yukawa couplings are
large; the coefficients $d'_k$, depending on gauge 
and Yukawa couplings, are of order unity.
Beyond the approximate solutions shown explicitly, the evolution equations 
have been solved numerically in the present analysis  to
two--loop order \cite{RGE2} and threshold effects have been
incorporated at the low scale \cite{bagger}.
The 2-loop effects as given in
Ref.~\cite{Degrassi:2001yf} have been included for
the neutral Higgs bosons and the $\mu$ parameter. 

\subsection{Gauge Coupling Unification}

Measurements of the gauge couplings at the electroweak scale
support very strongly the unification of the couplings at a scale
$M_U \simeq 2\times 10^{16}$~GeV \cite{r13A}.  
The precision, being at the per--cent level, is
surprisingly high after extrapolations over
fourteen orders of magnitude in the energy 
from the electroweak scale to the grand unification scale $M_U$. 
Conversely, the
electroweak mixing angle has been predicted in this approach at the
per--mille level. The evolution of the gauge couplings from 
low energy  to the GUT scale $M_U$ is carried out at two--loop accuracy. 
The gauge couplings $g_1$,
$g_2$, $g_3$ and the Yukawa couplings are calculated
in the $\overline{DR}$ scheme
by adopting the shifts given in Ref.\cite{bagger}.
These parameters are evolved to $M_U$ using 2--loop RGEs \cite{RGE2}. 
At 2-loop order the gauge couplings do not meet
exactly \cite{Weinberg:1980wa}, the
differences attributed to threshold effects at the unification
scale $M_U$ which leave us with an ambiguity in the definition of
$M_U$. In this report we define $M_U$ as the scale, {\it ad libitum}, 
where $\alpha_1 = \alpha_2$, denoted $\alpha_U$, in the RG evolution.
The non--zero 
difference $\alpha_3 - \alpha_U$ at this scale is then accounted
for by threshold effects
of particles with masses of order $M_U$. The quantitative evolution implies 
important constraints on the particle content at $M_U$
\cite{Ross:1992tz}.

\begin{figure*}
\setlength{\unitlength}{1mm}
\begin{center}
\begin{picture}(160,85)
\put(-25.5,-101.5){\mbox{\epsfig{figure=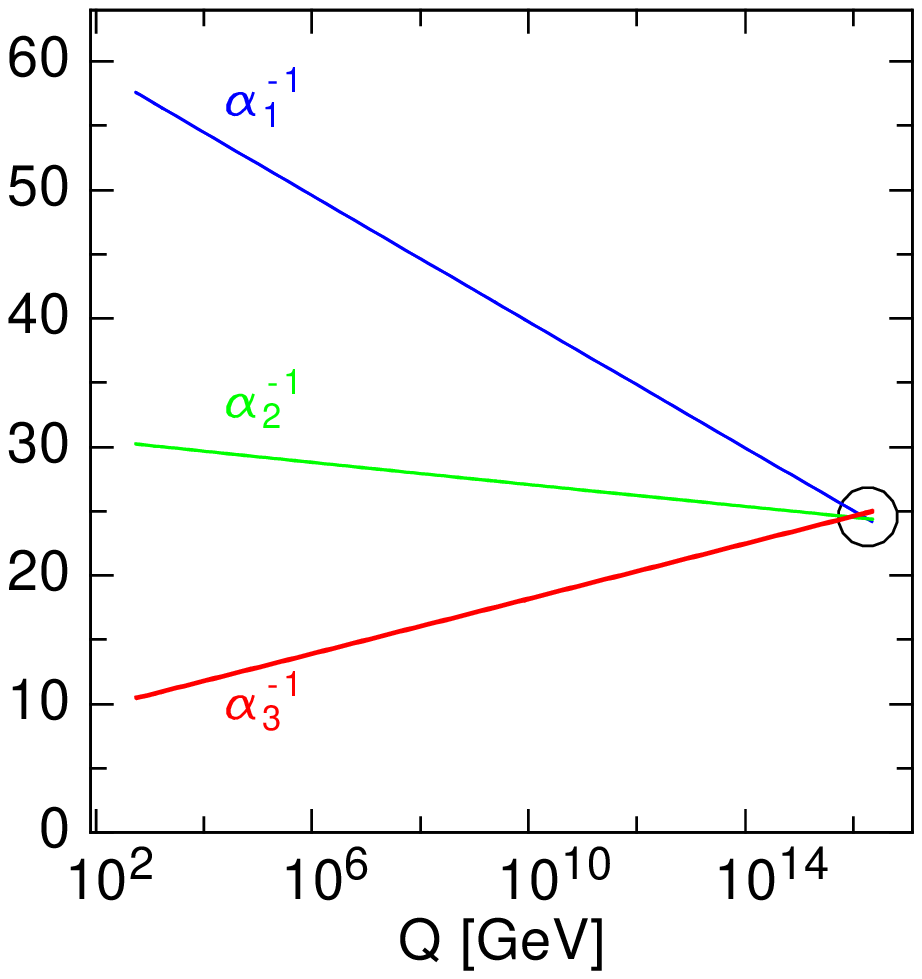,
                                   height=22.cm,width=16.cm}}}
\put(76,31.5){\mbox{\huge $\Rightarrow$}}
\put(60.5,-101.5){\mbox{\epsfig{figure=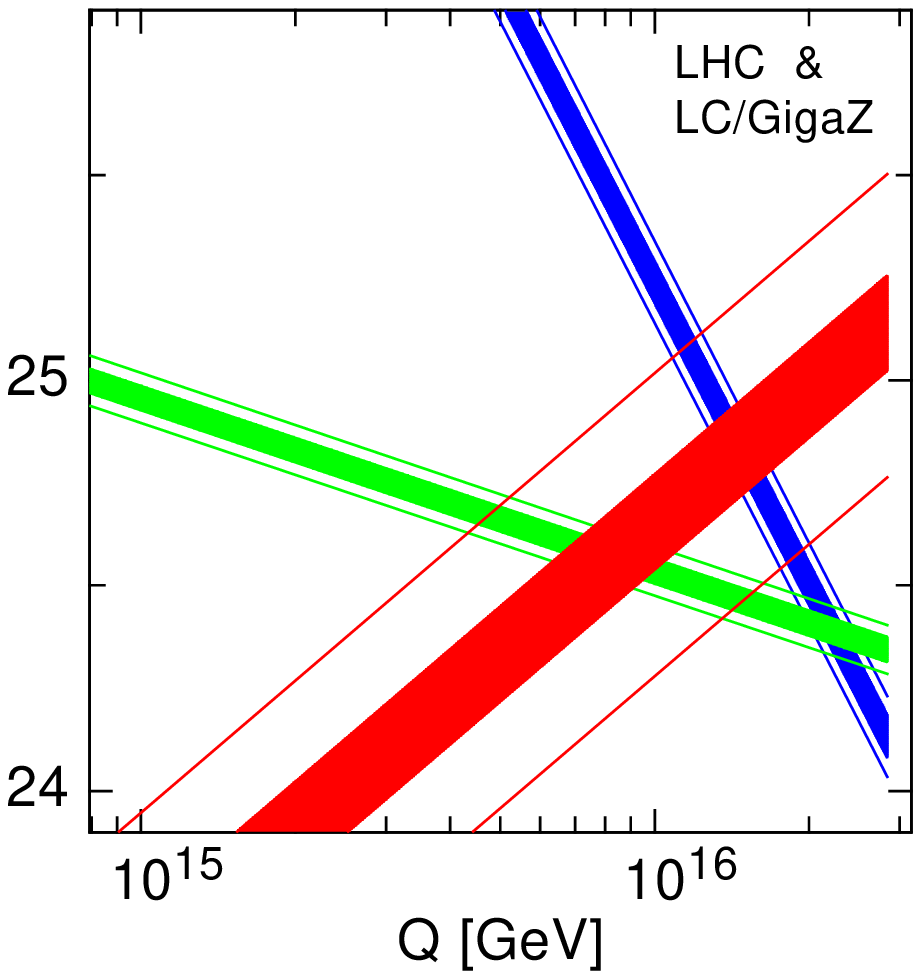,
                                   height=22.cm,width=16.cm}}}
\put(-3,80){\mbox{\bf a)}}
\put(84,80){\mbox{\bf b)}}
\end{picture}
\end{center}
\caption{{\it (a) Running of the inverse gauge couplings from low
  to high energies.
  (b) Expansion of the area around the unification point $M_U$ 
      defined by the meeting point of $\alpha_1$ with $\alpha_2$.
      The wide error bands are based on present data, and the spectrum
      of supersymmetric particles from LHC measurements within mSUGRA. 
      The narrow bands demonstrate the improvement expected by future GigaZ 
      analyses and the measurement of the complete spectrum at ``LHC+LC''.}}
\label{fig:gauge}
\end{figure*} 
\begin{table}
\begin{center}
\begin{tabular}{|c||c|c|}
\hline
 & Present/''LHC'' & GigaZ/''LHC+LC'' \\
\hline \hline
$M_U$ & $(2.36 \pm 0.06)\cdot 10^{16} \, \rm {GeV}$ & 
           $ (2.360 \pm 0.016) \cdot 10^{16} \, \rm {GeV}$ \\
$\alpha_U^{-1}$ & $  24.19 \pm 0.10 $ &  $ 24.19 \pm 0.05 $\\ \hline
$\alpha_3^{-1} - \alpha_U^{-1}$ & $0.97 \pm 0.45$ & $0.95 \pm 0.12$ \\ \hline
\end{tabular}
\end{center}
\caption{{\it Expected errors on $M_U$ and $\alpha_U$ for the mSUGRA 
 reference point, derived for the present level of accuracy and
 compared with expectations from GigaZ [supersymmetric spectrum as discussed
  in the text].  Also shown is the difference between
 $\alpha_3^{-1}$ and $\alpha_U^{-1}$ at the unification point $M_U$.}}
\label{tab:gauge}
\end{table}
Based on the set of low--energy gauge and Yukawa parameters 
$\{\alpha(m_Z)$, $\sin^2
\theta_W$, $\alpha_s(m_Z)$, $Y_t(m_Z)$, $Y_b(m_Z)$, $Y_\tau(m_Z)\}$
the evolution of the inverse couplings $\alpha_i^{-1}$ $[i=U(1)$, $SU(2)$,
$SU(3)]$ is depicted in Fig~\ref{fig:gauge}. The evolution is performed for
the mSUGRA reference point defined above. Unlike earlier analyses, the
low--energy thresholds of supersymmetric particles can be calculated
in this framework
exactly without reference to effective SUSY scales. The outer lines
in Fig.~\ref{fig:gauge}b
correspond to the present experimental accuracy of the gauge
couplings \cite{PDG}: 
$\Delta \{\alpha^{-1}(m_Z)$, $\sin^2\theta_W$, $\alpha_s(m_Z)\}$ 
$=\{ 0.03, 1.7 \cdot 10^{-4}, 3 \cdot 10^{-3} \}$, and the spectrum of
supersymmetric particles from LHC measurements complemented in the top-down
approach for mSUGRA. 
The full bands demonstrate the improvement for the absolute errors
$\{8 \cdot 10^{-3}, 10^{-5},10^{-3}  \}$ after operating GigaZ 
\cite{Monig:2001hy,Erler:2000jg} and inserting the complete spectrum
from ``LHC+LC'' measurements.  
The expected accuracies in $M_U$ and $\alpha_U$
are summarized in the values given in Tab.~\ref{tab:gauge}.
The gap between $\alpha_U$ and 
$\alpha_3$ is bridged by contributions from high scale physics.
Thus, for a typical set of SUSY parameters, the evolution of the gauge
couplings from low  to high scales leads to a precision of 1.5 per--cent
for the Grand Unification picture.

\subsection{Gaugino and Scalar Mass Parameters: Top-down Approach}

The structure of the fundamental supersymmetric theory is assumed,
in the top-down approach, to be defined
uniquely at a high scale. In mSUGRA the set of
parameters characterizing the specific form of the theory includes,
among others, the scalar masses $M_0$ and the gaugino masses $M_{1/2}$. 
These universal parameters are realized at the grand unification point $M_U$.
Evolving the parameters from the high scale down to the electroweak scale
leads to a comprehensive set of predictions for the masses, mixings and 
couplings of the physical particles. Precision measurements of these 
observables can be exploited to determine the high-scale parameters
 $M_0$, $M_{1/2}$, etc., and to perform consistency tests of the
underlying form of the theory. The small number of fundamental parameters,
altogether five in mSUGRA, gives rise to many correlations between a 
large number of experimental observables. They define a set of
{\it necessary consistency conditions} for the realization of the
specific fundamental theory in nature. \\

\noindent
{\it Interludium:}
In addition to the experimental errors, theoretical uncertainties must be
taken into account. They are generated by truncating the perturbation series
for the evolution of the fundamental parameters in the $\overline{DR}$ scheme
from the GUT scale to a low SUSY scale $\tilde M$ near the electroweak
scale, and for the relation between the parameters at this point to the
on-shell physical mass parameters, for instance. Truncating
these series in one- to two-loop approximations leads to a residual
$\tilde M$ dependence that would be absent from the exact solutions and may
therefore be interpreted as an estimate of the neglected higher-order 
effects.

\renewcommand{\arraystretch}{1.1}
\begin{table}
\begin{center}
\begin{tabular}{|c|c||c|c|}
\hline
Particle & $\Delta_{th}$~[GeV]
 & Particle & $\Delta_{th}$~[GeV] \\ 
\hline \hline
$\tilde \chi^+_1$ & 1.2 &  $\tilde q_R$ & 8.4\\
$\tilde \chi^+_2$ & 2.8 &  $\tilde q_L$ & 9.1\\ 
$\tilde \chi^0_1$ & 0.34 & $\tilde t_1$ & 4.4\\
$\tilde \chi^0_2$ & 1.1 & $\tilde t_2$ & 8.3 \\
$\tilde \chi^0_3$ & 0.6 &  $\tilde b_1$ & 7.4\\
$\tilde \chi^0_4$ & 0.3 &  $\tilde b_2$ & 8.2 \\ \hline
$\tilde e_R$ & 0.82 & $\tilde g$ & 1.2 \\ \cline{3-4}
 $\tilde e_L$ & 0.31 & $h^0$ & 1.2 \\ 
 $\tilde \nu_e$ & 0.24 & $H^0$ & 0.7 \\ 
 $\tilde \tau_1$ & 0.59 & $A^0$ & 0.7  \\
 $\tilde \tau_2$ & 0.30 & $H^+$ & 1.0 \\
 $\tilde \nu_\tau$ & 0.25 & & \\ \hline
\end{tabular}
\end{center}
\caption{{\it Theoretical errors of the SPS1a mass spectrum, 
calculated as difference between the minimal 
and the maximal value of the masses
if the scale $\tilde M$ is varied between 100 GeV and 1 TeV.}}
\label{tab:masserrorth}
\end{table}

\renewcommand{\arraystretch}{1.1}
\begin{table}
\begin{center}
\begin{tabular}{|c||cccccccc|}
\hline
 & $m_{ll}^{max}$  
 & $m_{llq}^{max}$ 
 & $m_{llq}^{min}$ 
 & $m_{lq}^{high}$ 
 & $m_{lq}^{low}$  
 & $m_{\tau\tau}^{max}$  
 & $m_{ll}^{max}(\tilde\chi^0_4)$ 
 & $m_{llb}^{min}$ \\ \hline \hline
\SPHENO & 80.64 & 454.0 & 216.8 & 397.2 & 325.6 & 83.4\; & 283.4 & 195.9 \\
$\Delta_{exp}$  
    & \hphantom{0}0.08 
    & \hphantom{00}4.5 
    & \hphantom{00}2.6 
    & \hphantom{00}3.9 
    & \hphantom{00}3.1 
    &  \hphantom{0}5.1 
    & \hphantom{00}2.3 
    & \hphantom{00}4.1 \\
$\Delta_{th}\;\,$ 
    & \hphantom{0}0.72 
    & \hphantom{00}8.1 
    & \hphantom{00}3.6 
    & \hphantom{00}7.7 
    & \hphantom{00}5.5 
    &  \hphantom{0}0.8 
    & \hphantom{00}0.7 
    & \hphantom{00}2.9 \\ \hline \hline
\end{tabular}\\[4mm]
\begin{tabular}{|c||cccccc|}
\hline 
 & $m_{\tilde q_R}-m_{\tilde\chi^0_1}$
 & $m_{\tilde l_L}-m_{\tilde\chi^0_1}$ 
 & $m_{\tilde g}-m_{\tilde b_1}$
 & $m_{\tilde g}-m_{\tilde b_2}$
 & $m_{\tilde g}-0.99\,m_{\tilde\chi^0_1}$
 & $m_{h^0}$ \\ \hline \hline
\SPHENO & 450.3 & 110.0 & 88.9 & 56.9 & 507.8 & 110.8 \\
$\Delta_{exp}$  
    & \phantom{0}10.9 
    & \phantom{00}1.6 
    & \phantom{0}1.8 
    & \phantom{0}2.6 
    & \phantom{00}6.4 
    & \phantom{000}0.25 \\
$\Delta_{th}\;\,$  
    & \phantom{00}8.1 
    & \phantom{000}0.23 
    & \phantom{0}6.8 
    &    \phantom{0}7.6 
    & \phantom{00}1.3 
    & \phantom{00}1.2 \\ \hline
\end{tabular}
\end{center}
\caption{{\it LHC observables assumed for SPS1a and their 
   experimental ($\Delta_{exp}$) and present theoretical ($\Delta_{th}$)  
   uncertainties. [All quantities in GeV].} 
\label{tab:LHCobs}}
\end{table}

We estimate these effects by varying $\tilde M$ between the electroweak
scale and 1 TeV. The theoretical uncertainties 
of the physical masses and  LHC observables derived in this way 
are listed in 
Tables~\ref{tab:masserrorth} and \ref{tab:LHCobs}, respectively. 
They are of similar size as the  differences found by comparing
the observables with different state-of-the-art 
codes for the spectra \cite{Allanach:2003jw}.
The comparison of the present theoretical
uncertainties with the experimental errors at LHC demonstrates that the
two quantities do match {\it cum grano salis} at the same size.
Since LC experiments will reduce the experimental errors roughly by an
order of magnitude, considerable theoretical efforts are needed 
in the future  to reduce $\Delta_{th}$ to a level that matches 
the expected experimental precision at  LC.  
Only then we can deepen our 
understanding of the underlying supersymmetric theory by tapping the full
experimental potential of ``LC'' and of the combined ``LHC+LC'' analyses. \\

\begin{figure*}
\setlength{\unitlength}{1mm}
\begin{center}
\begin{picture}(160,80)
\put(-10,-25){\mbox{\epsfig{figure=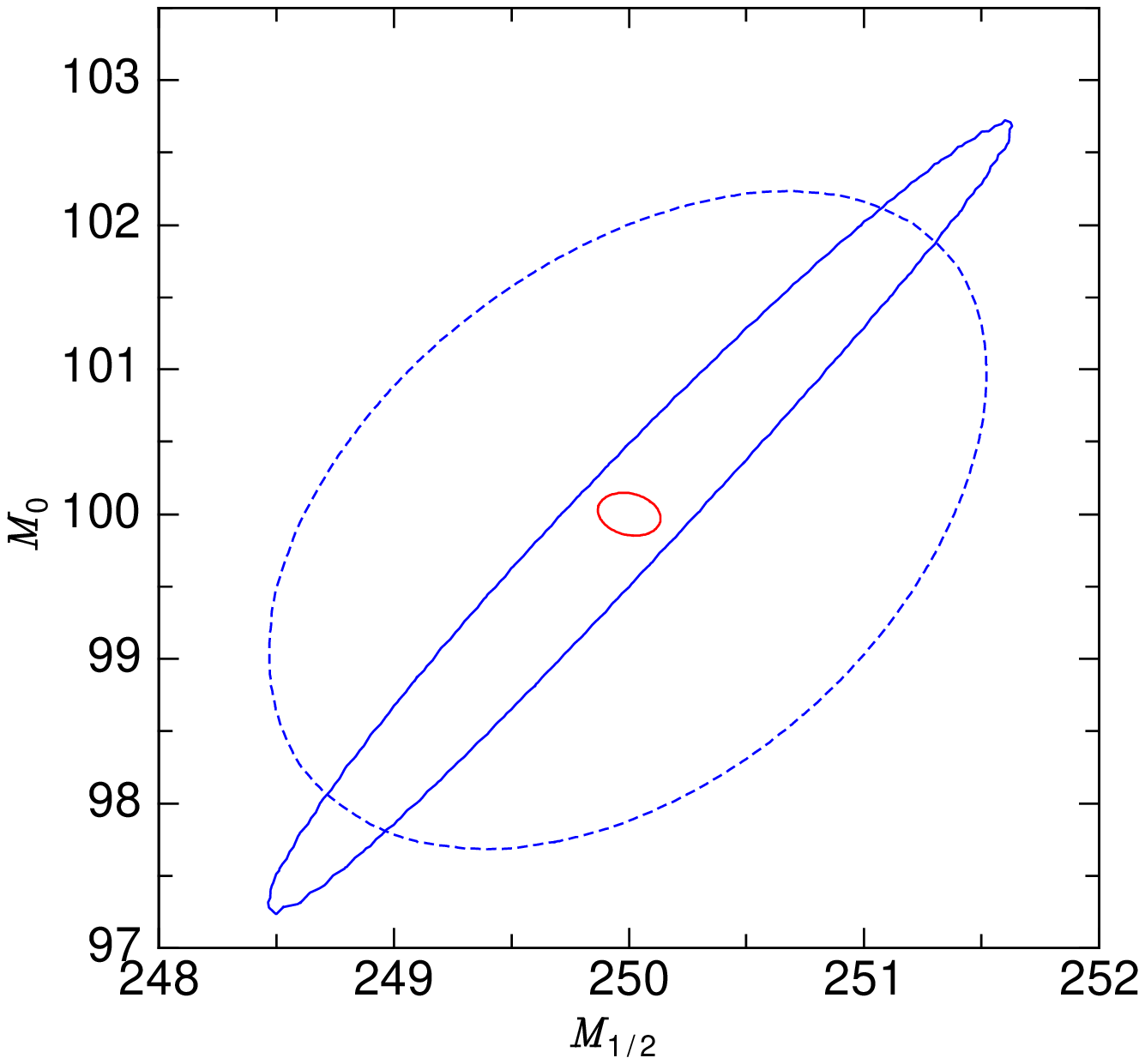,
                                   height=17.5cm,width=12.cm}}}
\put(74,-25){\mbox{\epsfig{figure=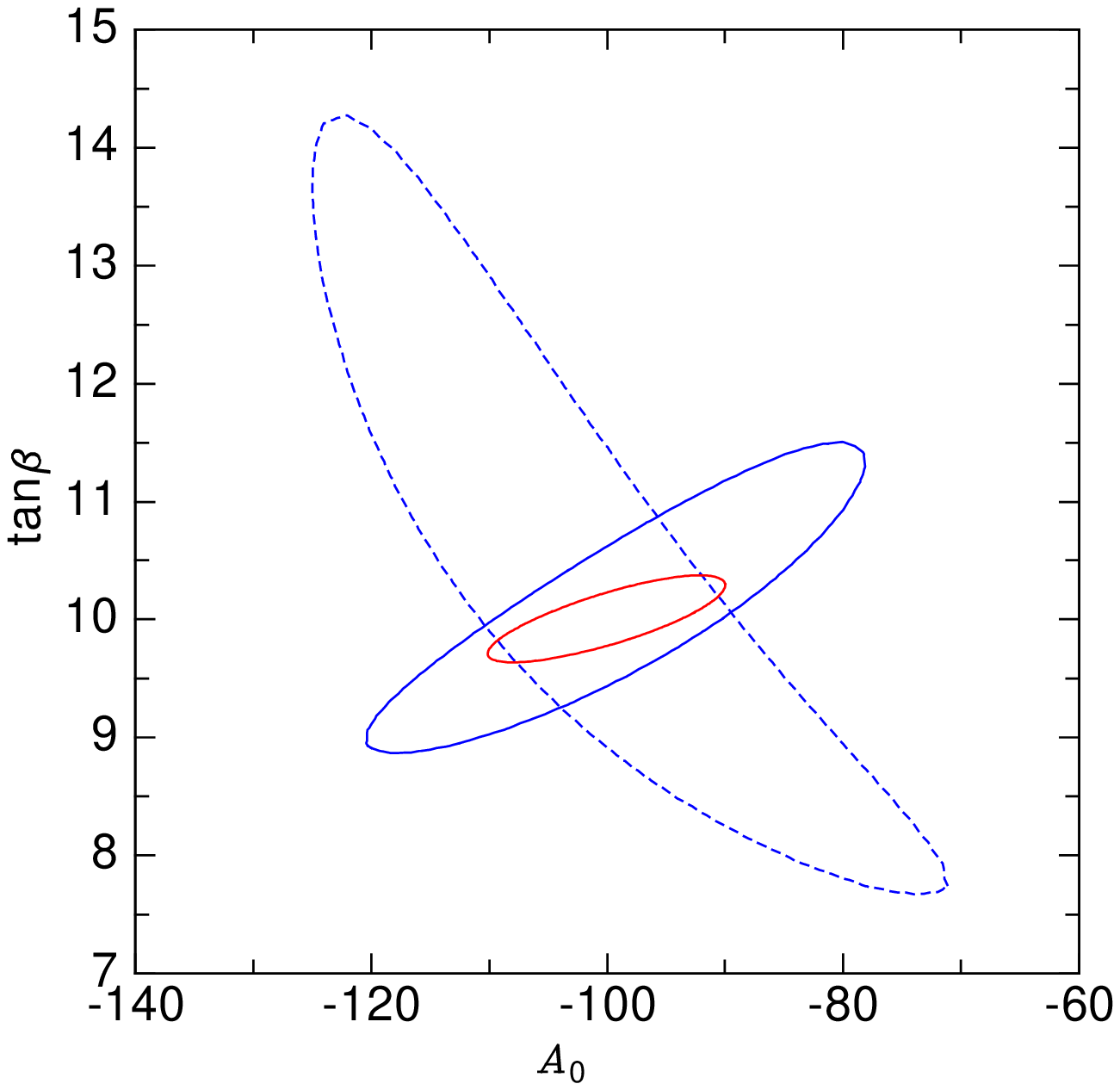,
                                   height=17.5cm,width=12.cm}}}
\put(-3,75){\mbox{\bf a)}}
\put(80,75){\mbox{\bf b)}}
\end{picture}
\end{center}
\caption{\it 1-$\sigma$ error ellipses for the mSUGRA parameters
  in the top-down approach [i.e. the contours of $\Delta\chi^2=4.7$ 
  in the $M_0-M_{1/2}$ and $\tan\beta-A_0$ planes with the respective 
  other parameters fixed to their best fit values, 
  c.f. Table~\ref{tab:topdownresults}].  
  The full blue ellipses are the results obtained from LHC measurements 
  alone while the red ones are for the combined ``LHC+LC'' analyses. 
  The dashed blue lines show the results for the ``LHC'' case including 
  today's theoretical uncertainties.}
\label{fig:topdown}
\end{figure*}

In the top-down approach, models of SUSY-breaking are tested by fitting 
their high-scale parameters to experimental data. 
The minimum $\chi^2$ of the fit gives a measure of the probability that 
the model is wrong. 
The results of such a fit of mSUGRA to anticipated ``LHC'', ``LC'', and ``LHC+LC'' 
measurements are shown in Table~\ref{tab:topdownresults} and 
Fig.~\ref{fig:topdown}. 
For the ``LHC'' case the observables in Table~\ref{tab:LHCobs} have been used, 
for the LC the masses in Table~\ref{tab:massesA} and for ``LHC+LC''
the complete information have been used. 
If mSUGRA is assumed to be the underlying supersymmetric theory, the 
universal parameters $M_{1/2}$ and $M_0$ can be determined at the LHC 
at the per--cent level. LC experiments and coherent ``LHC+LC'' analyzes 
improve the accuracy by an order of magnitude, thus allowing for much 
more powerful tests of the underlying supersymmetric theory. 
Table~\ref{tab:topdownresults} takes only experimental errors into account.
The accuracy of the present theoretical calculations matches the errors
of the ``LHC'' analysis and can thus be included in a meaningful way
in a combined experimental plus theoretical error analysis. 
Adding $\Delta_{th}$ and $\Delta_{exp}$ quadratically the
 errors of the ``LHC'' analysis increases to:
$\Delta M_{1/2}= 2.7$~GeV, $\Delta M_0= 2.9$~GeV, $\Delta A_0= 51$~GeV, and 
$\Delta\tan\beta= 5$. 
As argued above,  significant theoretical improvements by an order of
magnitude, i.e. ``the next loop'', are necessary to exploit fully
 the ``LC'' and ``LHC+LC'' potential.

The minimum $\chi^2$ of the fit to mSUGRA as in Table~\ref{tab:topdownresults}
 is indeed small, 
$\chi^2_{min}/n.d.o.f. \leq 0.34$ for ``LHC'', ``LC'', as well as ``LHC+LC''. 
When fitting instead mGMSB model parameters as an alternative 
to the same data, 
we would obtain $\chi^2_{min}/14\,d.o.f. = 68$ from LHC data alone. 
Such a result would clearly disfavour this model.

\begin{table}
\begin{center}
\begin{tabular}{|c||c|c||c|}
\hline
             & ``LHC''        & ``LC'' & ``LHC+LC'' \\ \hline \hline
$M_{1/2}$   & $250.0 \pm 2.1$ &  $250.0\pm 0.4$ & $250.0 \pm 0.2$  \\
$M_0$       & $100.0 \pm 2.8$ &  $100.0\pm 0.2$ & $100.0 \pm 0.2$  \\
$A_0$       & $-100.0\pm 34$  & $-100.0\pm 27$  & $-100.0\pm 14$   \\
$\tan\beta$ & $ 10.0 \pm 1.8$ &  $10.0 \pm 0.6$ & $10.0 \pm 0.4$  \\
\hline
\end{tabular}
\end{center}
\caption{{\it Results for the high scale parameters in the
   top-down approach including the experimental errors.}}
\label{tab:topdownresults}
\end{table}

\subsection{Gaugino and Scalar Mass Parameters: Bottom-up Approach}

In the bottom-up approach the fundamental supersymmetric theory
is reconstructed at the high scale from the available {\it corpus} of
experimental data without any theoretical prejudice. This approach exploits
the experimental information to the maximum extent possible and reflects an
undistorted picture of our understanding of the basic theory.
 
\begin{figure*}
\setlength{\unitlength}{1mm}
\begin{center}
\begin{picture}(160,170)
\put(-4,0){\mbox{\epsfig{figure=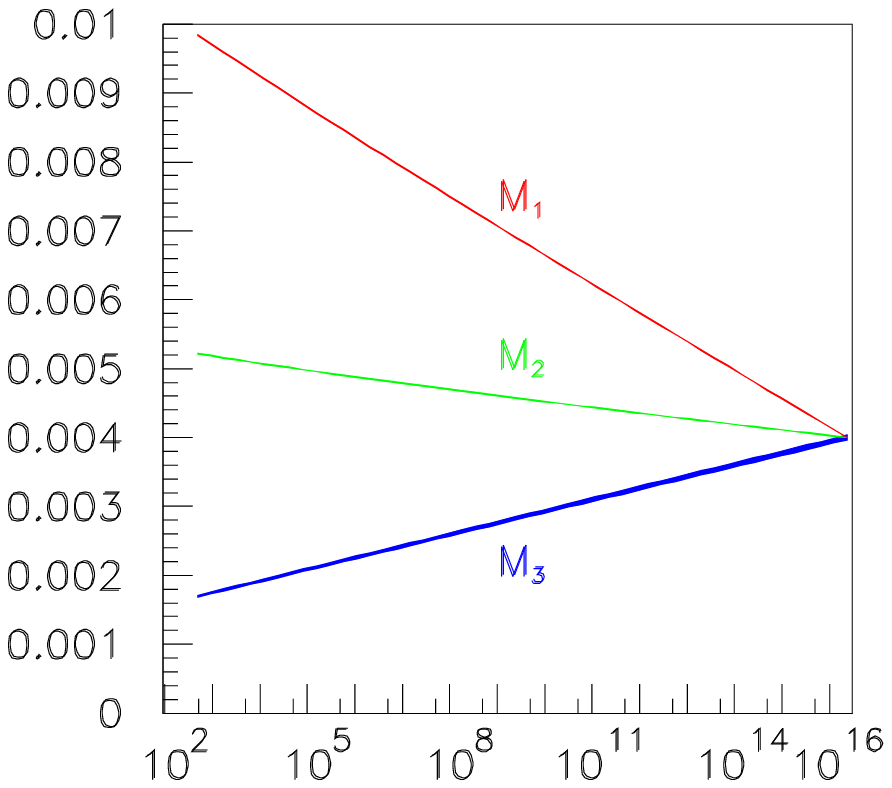,height=17cm,width=18cm}}}
\put(84.5,83.5){\mbox{\epsfig{figure=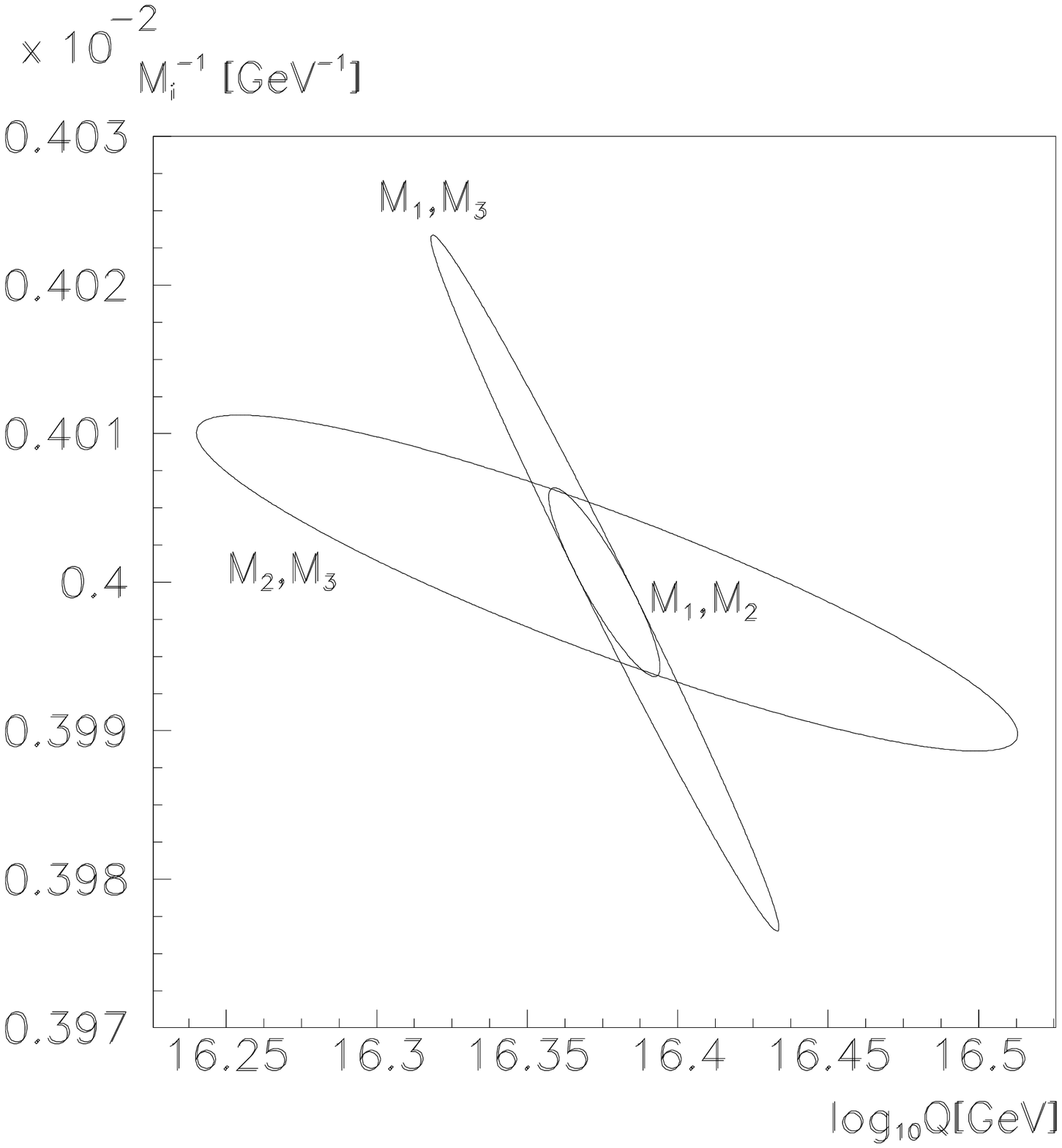
                              ,height=7.95cm,width=8.3cm}}}
\put(-4,-86){\mbox{\epsfig{figure=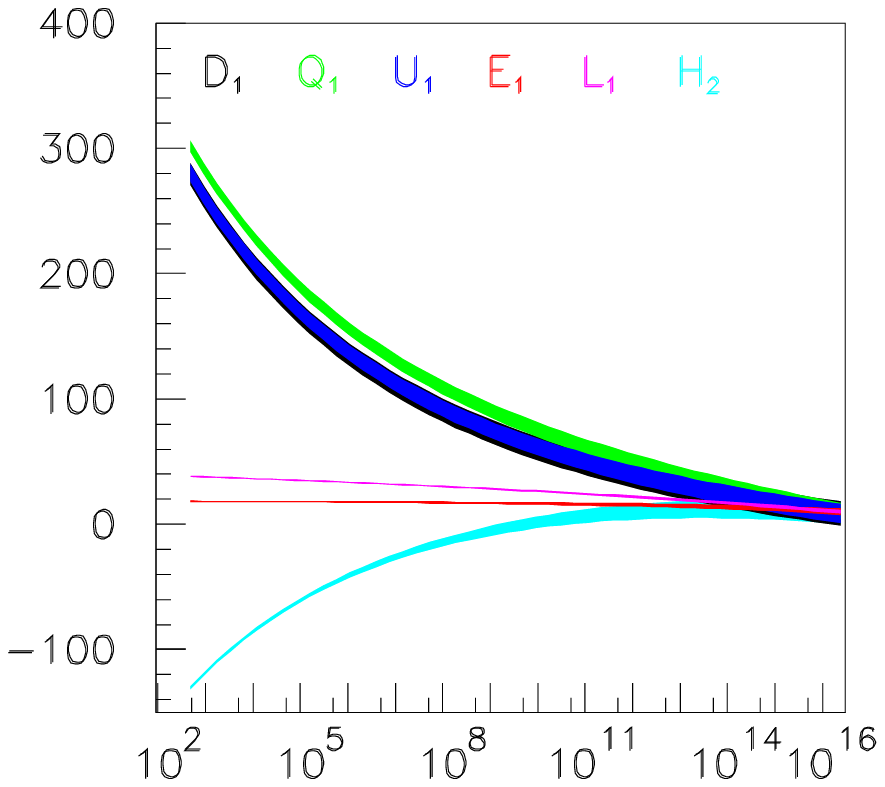,height=17cm,width=18cm}}}
\put(78,-86){\mbox{\epsfig{figure=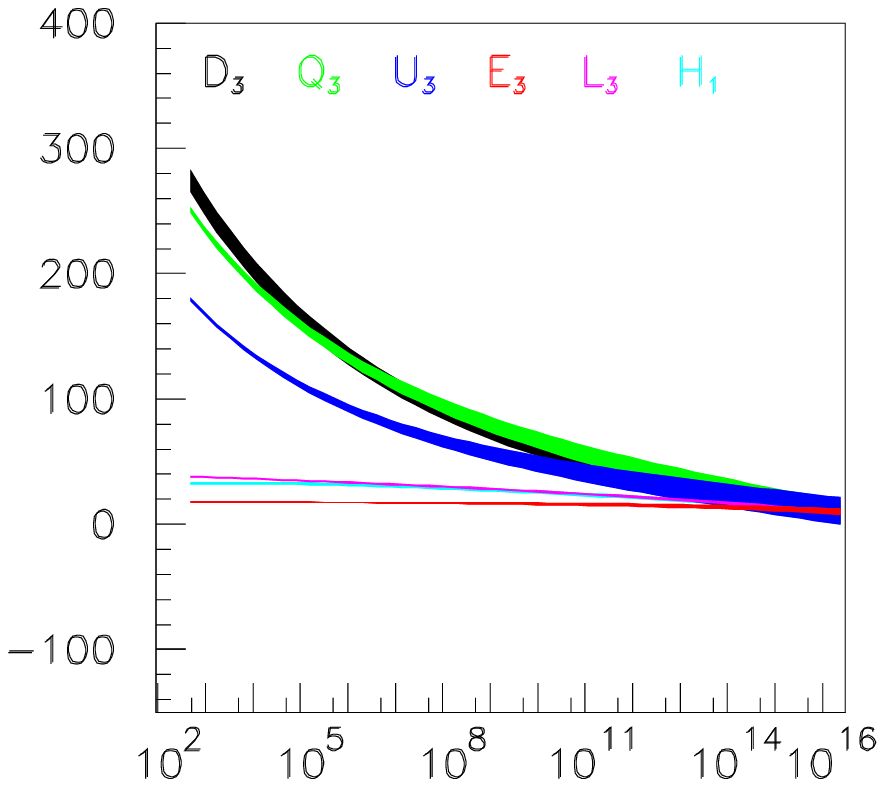,height=17cm,width=18cm}}}
\put(-1,158){\mbox{\bf (a)}}
\put(10,156){\mbox{$1/M_i$~[GeV$^{-1}$]}}
\put(60,82){\mbox{$Q$~[GeV]}}
\put(-1,73){\mbox{\bf (b)}}
\put(14,71){\mbox{$M^2_{\tilde j}$~[$10^3$ GeV$^2$]}}
\put(65,-3){\mbox{$Q$~[GeV]}}
\put(80,159){\mbox{
}}
\put(80,73){\mbox{
}}
\put(95,71){\mbox{$M^2_{\tilde j}$~[$10^3$ GeV$^2$]}}
\put(147,-3){\mbox{$Q$~[GeV]}}
\end{picture}
\end{center}
\caption{{\it  Evolution, from low to high scales, (a) of 
the gaugino mass parameters
for ``LHC+LC'' analyses
and the corresponding error ellipses of the universal GUT values;
(b) left: of the first--generation sfermion mass parameters 
(second generation, {\it dito}) and  
the Higgs mass parameter $M^2_{H_2}$; right: of the
third--generation sfermion mass parameters and
the Higgs mass parameter $M^1_{H_2}$.}   
}
\label{fig:sugra_LHC}
\end{figure*}

At the present level of preparation in the ``LHC'' and ``LC'' sectors, such
a program can only be carried out in coherent ``LHC+LC'' analyses  while
the separate information from either machine proves insufficient.
The results for the evolution of the mass parameters from the electroweak
scale to the GUT
scale $M_U$ are shown in Fig.~\ref{fig:sugra_LHC}.  

On the left of Fig.~\ref{fig:sugra_LHC}a
the evolution is presented for the 
gaugino parameters $M^{-1}_i$, which clearly
is under excellent control for the coherent ``LHC+LC'' analyses, 
while ``LHC'' [and ``LC''] measurements alone are insufficient for the 
model-independent reconstruction
of the parameters and the test of universality
in the $SU(3) \times SU(2) \times U(1)$ group space.
The error ellipse for the unification of the gaugino masses 
in the final analysis is depicted on the right of 
Fig.~\ref{fig:sugra_LHC}a. Technical details of the ``LHC+LC'' analysis
can be found in Ref.~\cite{r1}. 

In the same way the evolution of the scalar mass parameters can be
studied, presented in Figs.~\ref{fig:sugra_LHC}b 
separately for the first/second
and the third generation in ``LHC+LC'' analyses.
Compared with the slepton parameters, the accuracy deteriorates
for the squark parameters, 
and for the Higgs mass parameter
$M^2_{H_2}$.
The origin of the differences between the errors for slepton and 
squark/Higgs mass parameters can be traced back to the numerical 
size of the coefficients
in Eqs.~(\ref{eq:squark}). Typical examples, 
evaluated at $Q=500$~GeV, read as follows \cite{r1}:
\begin{eqnarray}
M^2_{\tilde L_{1}} &\simeq& M_0^{2} + 0.47 M^2_{1/2} \\
M^2_{\tilde Q_{1}} &\simeq& M_0^{2} + 5.0 M^2_{1/2}  \\
M^2_{\tilde H_2} &\simeq&  -0.03 M_0^{2} - 1.34 M^2_{1/2}
           + 1.5 A_0 M_{1/2} + 0.6 A^2_0 \\
|\mu|^2 &\simeq& 0.03 M_0^{2} + 1.17 M^2_{1/2}
           - 2.0 A_0 M_{1/2} - 0.9 A^2_0
\end{eqnarray}
While the coefficients for the sleptons are of order unity, 
the coefficients $c_j$
for the squarks grow very large,  $c_j \simeq 5.0$, so that small errors
in $M^2_{1/2}$ are magnified by nearly an order of magnitude in the solution
for $M_0$. By close inspection of Eqs.(\ref{eq:squark}) for the Higgs mass
parameter it turns out that 
the formally leading 
$M^2_0$ part is nearly cancelled by the $M^2_0$ part
of $c'_{j,\beta} \Delta M_\beta^2$. Inverting Eqs.(\ref{eq:squark}) for
$M^2_0$ therefore gives rise to large errors in the Higgs case.
Extracting the trilinear parameters $A_k$ is difficult and
more refined analyses
based on sfermion cross sections and Higgs and/or sfermion decays are
necessary to determine these parameters accurately.

A representative set of the final mass values and the associated errors,
after evolution from the electroweak scale to $M_U$, are presented
in Table~\ref{tab:evolved_params}.
It appears that the joint ``LHC+LC''analysis generates
a comprehensive and detailed picture of the fundamental SUSY parameters 
at the GUT/PL scale. Significant improvements however would be welcome
in the squark sector where reduced experimental errors would refine the
picture greatly.

\renewcommand{\arraystretch}{1.1}
\begin{table}
\begin{center}
\begin{tabular}{|c||c|c|}
\hline
               & Parameter, ideal & ``LHC+LC'' errors     \\
\hline\hline
 $M_1$        &  $250.$           & 0.15   \\
 $M_2$        &  {\it ditto}     & 0.25   \\
 $M_3$        &                   & 2.3    \\
\hline\hline  
 $M_{L_1}$  &$100.$     & 6.  \\
 $M_{E_1}$  &{\it ditto}          & 12.  \\
 $M_{Q_1}$  &                   & 23.  \\
 $M_{U_1}$  &                   & 48.  \\
 $M_{L_3}$  &                   & 7.  \\
 $M_{E_3}$  &                   & 14. \\
 $M_{Q_3}$  &                   & 37. \\
 $M_{U_3}$  &                   & 58.  \\
\hline
$M_{H_1} $  &{\it ditto}          & 8.  \\
$M_{H_2} $  &                   & 41.  \\
\hline \hline
$A_t       $  &  $-100.$          & 40.                 \\
\hline
\end{tabular}\\
\end{center}
\caption {
{\it Values of the SUSY Lagrange mass parameters after extrapolation 
to the unification scale where gaugino and scalar mass parameters are 
universal in mSUGRA [mass units in {\rm GeV}].} 
} 
\label{tab:evolved_params}
\end{table}

\section{Summary}
We have shown in this brief report that in supersymmetric theories stable
extrapolations can be performed from the electroweak scale
to the grand unification scale, close to the Planck scale.
This feature has been demonstrated compellingly in the evolution
of the three gauge couplings and of the soft supersymmetry breaking
parameters, which approach universal values at the GUT scale in minimal 
supergravity.  As a detailed scenario we have adopted the Snowmass reference
point SPS1a. It turns out that the information on the mSUGRA
parameters at the GUT scale from pure ``LHC'' analyses
is too limited to allow for the reconstruction of the high-scale
theory in a model-independent way. 
The coherent ``LHC+LC'' analyses however in which the measurements 
of SUSY particle
properties at LHC and LC mutually improve each other, result in a
comprehensive and detailed picture 
of the supersymmetric particle system. In particular, the gaugino sector
and the non-colored scalar sector are under excellent control.
 
\begin{table}
\begin{center}
\begin{tabular}{|c||c|c|}
\hline
                &  Parameter, ideal    & Experimental error \\ 
\hline\hline
$M_U$           & $2.36\cdot 10^{16}$ &  $2.2  \cdot 10^{14}$       \\
$\alpha_U^{-1}$ &   24.19          &     0.05     \\ \hline
$M_\frac{1}{2}$ & 250.             & 0.2     \\
$M_0$           & 100.             & 0.2     \\
$A_0$           & -100.            & 14      \\  
\hline
$\mu$           & 357.4            & 0.4     \\
\hline
$\tan\beta $    &  10.             & 0.4      \\  
\hline
\end{tabular}
\end{center}
\caption[]{\it Comparison of the ideal parameters with the
experimental expectations 
in the combined ``LHC+LC'' analyses
for the particular mSUGRA reference 
point adopted in this report [units in {\rm GeV}].} 
\label{tab:univ_params}
\end{table}
 
Though mSUGRA has been chosen as a specific example, the methodology can
equally well be applied to left-right symmetric theories and to
superstring theories.  The analyses offer the exciting opportunity to 
determine intermediate scales in left-right symmetric theories and to
measure effective string-theory parameters near the Planck scale.

Thus, a thorough analysis of the mechanism of
supersymmetry breaking and the reconstruction of the
fundamental supersymmetric theory at the grand unification scale
has been shown possible in the high-precision high-energy experiments at
LHC and LC.  This point has been highlighted by performing a global
mSUGRA fit of the universal parameters, c.f. Tab.~\ref{tab:univ_params}
Accuracies at the level of per-cent to per-mille can be
reached, allowing us to reconstruct the structure of nature
at scales where gravity is linked with particle physics.

\subsection*{Acknowledgments}
W.P.~is supported by the 'Erwin Schr\"odinger fellowship No.~J2272' 
of the `Fonds zur F\"orderung der wissenschaftlichen Forschung' of 
Austria and partly by the Swiss `Nationalfonds'.

%%%%%%%%%%%%%%%%%%%%%%%%%%%%%%%%%%%%%%%%%%%%%%%%%%%%%%%%%%%%%%
%%%%%%%%%%%%%%%%%%%%%%%%%%%%%%%%%%%%%%%%%%%%%%%%%%%%%%%%%%%%%%

\end{document}